\documentclass[letter]{elsarticle}

\usepackage{lineno,hyperref}
\modulolinenumbers[5]

\usepackage{graphicx}        

\newcommand{\be}{\begin{equation}}
\newcommand{\ee}{\end{equation}}
\newcommand{\bra}{\langle}
\newcommand{\ket}{\rangle}
\newcommand{\bea}{\begin{eqnarray}}
\newcommand{\eea}{\end{eqnarray}}

\journal{Journal of \LaTeX\ Templates}

\biboptions{authoryear}

\begin{document}

\begin{frontmatter}

\title{Market efficiency, liquidity, and multifractality of Bitcoin: 
A dynamic study}

\author{Tetsuya Takaishi\corref{mycorrespondingauthor}}
\cortext[mycorrespondingauthor]{Corresponding author}
\address{Hiroshima University of Economics, Hiroshima 731-0192 JAPAN}
\ead{tt-taka@hue.ac.jp}

\author{Takanori Adachi\corref{}}
\address{Tokyo Metropolitan University, Tokyo 100-0005 JAPAN}

\begin{abstract}

This letter investigates the dynamic relationship between market efficiency, liquidity, and multifractality of Bitcoin.
We find that before 2013 liquidity is low and the Hurst exponent is less than 0.5,
indicating that the Bitcoin time series is anti-persistent.
After 2013, as liquidity increased, the Hurst exponent 
 rose to approximately 0.5, improving market efficiency.
For several periods, however, the Hurst exponent was found to be significantly less than 0.5, making 
the time series anti-persistent during those periods.
We also investigate the multifractal degree of the Bitcoin time series using 
the generalized Hurst exponent and find that the multifractal degree is related 
to market efficiency in a non-linear manner.

\end{abstract}


\end{frontmatter}



\section{Introduction}

Bitcoin is the first practical cryptocurrency 
based on the blockchain technology required 
to maintain decentralized systems devised by \citet{Nakamoto2008}.
Since then, thousands of cryptocurencies have been proposed and created \citep{Coinmarketcap}. 
Bitcoin remains the most popular cryptocurrency and has the highest capitalization for cryptocurrencies (as of January 2019)\citep{Coinmarketcap}.

As Bitcoin has become widely recognized as payment medium,
it has naturally attracted the interest of researchers.
In recent years, a large number of studies focus on 
price properties and financial aspects of Bitcoin, such as
hedging capabilities \citep{dyhrberg2016hedging}, bubbles \citep{cheah2015speculative},
volatility analysis \citep{dyhrberg2016bitcoin,katsiampa2017volatility},
multifractality \citep{takaishi2018}, adaptive market hypothesis  \citep{khuntia2018adaptive},
transaction activity \citep{koutmos2018bitcoin}, Taylor  effect \citep{takaishi2018taylor}, 
investor attention \citep{urquhart2018causes}, and long memory effects\citep{phillip2018long}.

One of the most active research areas in this emerging field is the market efficiency of Bitcoin.
\citet{urquhart2016inefficiency} was the first study to test the market efficiency of Bitcoin. For the sample covering the period 2010-2016, that study finds 
that the Hurst exponent of the Bitcoin time series is less than 0.5,
indicating that the time series is anti-persistent. From this observation,
it is concluded that the Bitcoin time series is consistent with inefficient markets but may be in the process of moving toward an efficient market. 
\citet{bariviera2017inefficiency}  also calculates the Hurst exponent dynamically
and claims that, from 2011 until 2014, the Bitcoin time series has a Hurst exponent larger than 0.5,
i.e. the time series is persistent. Moreover, after 2011 the time series becomes compatible with white noise.
This conclusion squarely disagrees with Urquhart's results showing the anti-persistence of the time series. 

In addition to those two studies, several other papers have investigated the topic of market efficiency. 
However the conclusions are quite divergent.
\citet{tiwari2018informational} studies Bitcoin from 2010 to 2017 and reports that the Bitcoin market is efficient.
\citet{caporale2018persistence} examines the cryptocurrency market from 2013 to 2017 and 
finds that the Bitcoin market is inefficient, not as a consequence of anti-persistence, but as a consequence of persistence.
\citet{jiang2018time} and \citet{al2018efficiency} also report inefficiency caused by persistence in the time series for returns. 
\citet{alvarez2018long} reports that the Bitcoin market from 2013 to 2017 is not uniformly efficient
since anti-persistence of price returns appears cyclically. 
Based on the efficiency index, \citet{kristoufek2018bitcoin} finds strong evidence that the Bitcoin markets remained mostly inefficient
between 2010 and 2017. By analyzing one-minute returns from 2014 to 2016, \citet{takaishi2018} finds anti-persistence 
for high-frequency Bitcoin time series. \citet{sensoy2018inefficiency} and \citet{zargar2019informational} also report inefficiency of Bitcoin for high-frequency returns.

The aim of this letter is to examine the market efficiency of Bitcoin
and to contribute to the discussion on the nature of Bitcoin.
In particular, we analyze the dynamic properties of market efficiency and liquidity.
Recently, \citet{wei2018liquidity} examined the liquidity of cryptocurrencies and 
found that illiquid cryptocurrencies display inefficiency.
Thus, liquidity may have important implications in terms of market efficiency.
In addition, we calculate the generalized Hurst exponent, 
which characterizes  the multifractal nature of the time series.
The multifractality or generalized Hurst exponent of Bitcoin has also been 
addressed in the literature\citep{takaishi2018,jiang2018time,al2018efficiency,el2018bitcoin}.
Since Gaussian random time series show no multifractality, 
it has been suggested that the multifractal degree may be related to
the degree to which a time series deviates from efficiency.
Hence, we examine the relationship between  the multifractal degree and market efficiency.

This letter is organized as follows. Section 2 describes the data and methodology.
Section 3 presents the empirical results. Section 4 concludes.

\section{Data and Methodology}

In this study, we use Bitcoin Tick data (in dollars) traded on Bitstamp\footnote{Since we find no 
trading data from January 4, 2015 to January 9, 2015 due to the hacking incident to Bitstamp, 
we patch the missing data with the data from Bitfinex 
at Bitcoincharts.}
from September 14, 2012 to 
October 31, 2018 and downloaded from Bitcoincharts\footnote{http://api.bitcoincharts.com/v1/csv/}. 
Additionally, we use Bitcoin prices from two sources: 
daily Bitcoin price data (in Japanese Yen) from Coincheck from November 1, 2014 to October 31, 2018 
downloaded from Bitcoincharts and 
the daily Bitcoin price index created by Coindesk\footnote{http://www.coindesk.com/} from October 1, 2013 to
October 31, 2018.

Let $p_{t_n}; t_n= n\Delta t;  n=1,2,...,N$ be the time series of asset prices
 with sampling period $\Delta t$.
We define the return $r_{t_n}$ by the logarithmic price difference, namely,
\be
r_{t_{n+1}}=\log p_{t_{n+1}} -\log p_{t_{n}}.
\ee
In this study, we consider returns for $\Delta t=1,6,12,$ and $24$ hours.

To estimate the generalized Hurst exponent, we use the multifractal detrended fluctuation analysis (MF-DFA) method, 
which may be applied to non-stationary time series \citep{kantelhardt2002multifractal}. 
The MF--DFA is described as follows.

(i) Determine the profile $Y(i)$,
\be
Y(i)=\sum_{j=1}^i (r_{t_j}- \bra r \ket),
\ee
where $\bra r \ket$ stands for the average of returns.

(ii) Divide the profile $Y(i)$ into $N_s$ non-overlapping segments of equal length $s$, where $N_s \equiv {int} (N/s)$.
Since the length of the time series is not always a multiple of $s$, a short time period at the end of the profile may remain.
To utilize this part, the same procedure is repeated starting from the end of the profile.
Therefore, in total $2N_s$ segments are obtained.

(iii) Calculate the variance
\be
F^2(\nu,s)=\frac1s\sum_{i=1}^s (Y[(\nu-1)s+i] -P_\nu (i))^2,
\ee
for each segment $\nu, \nu=1,...,N_s$ and
\be
F^2(\nu,s)=\frac1s\sum_{i=1}^s (Y[N-(\nu-N_s)s+i] -P_\nu (i))^2,
\ee
for each segment $\nu, \nu=N_s+1,...,2N_s$.
Here, $P_\nu (i)$ is the fitting polynomial to remove the local trend in segment $\nu$;
we use a cubic order polynomial.

(iv) Average over all segments and obtain the $q$th order fluctuation function
\be
F_q(s)=\left\{\frac1{2N_s} \sum_{\nu=1}^{2N_s} (F^2(\nu,s))^{q/2}\right\}^{1/q}.
\label{eq:FL}
\ee

(v) Determine the scaling behavior of the fluctuation function.
If the time series $r_{t_i}$ are long-range power law correlated,
$F_q(s)$ is expected to be the following functional form for large $s$.
\be
F_q(s) \sim s^{h(q)}.
\label{eq:asympto}
\ee
The scaling exponent $h(q)$ is called the generalized Hurst exponent.

To describe the dynamic behavior of $h(q)$, we calculate $F_q(s)$ in a rolling window  with window-size equal to one year.
In this study, we calculate $h(q)$ by varying $q$ from $q_{min}=-25$ to $q_{max}=25$.
$h(2)$ corresponds to the usual Hurst exponent, and for $h(2)<0.5$ ( $h(2)>0.5$ ) the time series is classified as anti-persistent (persistent).
When $h(q)$ varies with $q$, the time series is said to be multifractal.
Conversely, when $h(q)$ is constant with varying $q$,  the time series is mono-fractal.
The Gaussian random walk is mono-fractal and  we expect $h(q)=0.5$ for any $q$.

Since the Gaussian time series shows no multifractality, 
the degree of multifractality is expected to relate with  some degree of deviation
from the Gaussian random walk, or market inefficiency. The relationship between the multifractal degree and stock market efficiency 
has been discussed in \citet{zunino2008multifractal}. 
Following \citet{zunino2008multifractal}, we define the degree of multifractality $\Delta h$ by
\be
\Delta h = h(q_{min})-h(q_{max}),
\ee
For this study, we take $q_{min}=-25$ and $q_{max}=25$, 
and investigate a relationship between multifractality degree $\Delta h$
and market efficiency as measured by the Hurst exponent $h(2)$.

To examine the relationship between market efficiency and liquidity, we
follow \citet{wei2018liquidity} and use the Amihud illiquidity measure as  proxy for illiquidity.
The Amihud illiquidity (ILLIQ) is defined as
\be
ILLIQ =\frac1D_T \sum _{t=1}^{D_T}\frac{|R_t|}{p_t V_t},
\ee
where $D_T$ is the number of days in the rolling window ( in this study, 1 year ), $R_t$ is the daily return on day $t$,
$V_t$ is the daily volume on day $t$, and $p_t$ is the daily closing price on day $t$.

\section{Empirical Results}

First, we compare the values of $h(2)$ calculated from the time series of returns sampled at the various frequencies. 
In Figure 1, we show the dynamical evolution of $h(2)$ for returns at $\Delta t=1, 6, 12,$ and $24$ hours.
The bottom of the figure also shows $h(2)$ calculated with daily (24-hour) returns from Coinchek and Coindesk.
From the figure it is clear that there is no significant difference between the three data sources for $h(2)$ based on daily returns.

All returns at the various sampling periods 
exhibit similar dynamic behavior, namely, before 2013, $h(2)$ is substantially below 0.5 
indicating that the time series is anti-persistent and,
after 2013, $h(2)$ increases to approximately $h(2)=0.5$ and fluctuates around this value. 
Even after $h(2)$ reaches the value of $0.5$, it is also observed that for some periods $h(2)$ falls significantly below 0.5. 
This is especially true during the period from  mid-2017 to  mid=2018, where $h(2)$ falls to  approximately 0.4. 
Since we find no significant difference between the results at the various sampling periods,
hereafter we focus on the results at $\Delta t= 24$ hours.

To check whether $h(2)$  is related to prices and/or returns, in Figure 2 we plot  these three variables together, and   observe no special relationship between $h(2)$ and returns.
On the other hand, $h(2)$ in the period from  mid-2017 to  mid-2018 appears to be related to price behavior.
In fact, during this period price increases drastically exhibited bubble-like behavior. Indeed, from the beginning of 2018, the price falls rapidly like a speculative bubble bursting.
Thus, for the period from  mid-2017 to  mid-2018 during which Bitcoin was unstable, 
the time series exhibited anti-persistence. 
For other periods, where $h(2)$ is below 0.5, we find no clear relationship in this graphic scale.

Second, we focus on the effect of liquidity. 
Figure 2 shows the Amihud illiquidity measure (ILLIQ) as a function of time.
We find that before 2013, ILLIQ is very high, typically greater than $10^{-6}$, and then it decreases with time.
Figure 3 shows a scatterplot of $h(2)$ versus ILLIQ.
We find that as ILLIQ decreases, $h(2)$ approaches 0.5.
When ILLIQ is high enough, for instance for $ILLIQ > 10^{-6}$, $h(2)$ is significantly less than 0.5, 
thus indicating that for illiquid periods the Bitcoin time series is anti-persistent.

Finally,  
we examine the relationship between market efficiency and the multifractal degree.
Figure 5 shows the multifractal degree $\Delta h$ as a function of time. The figure shows that $\Delta h$ gradually decreases with time,
except for the period from mid-2017 to mid-2018 in which the anti-persistent behavior was observed.
The gradual decrease of $\Delta h$  may indicate that the deviation from the Gaussian random 
walk also decreases.
Figure 6 shows  scatterplots of  $h(2)$ and the multifractal degree $\Delta h$.
While circle symbols indicate the data from the period from 2012 to 2013, diamond symbols are those from 2014 to 2018.
Notably we find  a non-linear relationship between market efficiency and multifractal degree. 
In fact, for  $\Delta h \leq 0.75$, $h(2)$ stays around 0.5, 
which means that the time series is efficient and no strong correlation between $h(2)$ and $\Delta h$ is present. 
However, as $\Delta h$ increases from 0.75, $h(2)$ declines rapidly from 0.5.
Thus, when $\Delta h$ is large, namely, larger than the threshold located at approximately 0.75, 
the time series becomes inefficient, as demonstrated by the observed anti-persistent behavior.
We also recognize that most of the data from the period from 2012 to 2013, in which the liquidity is low, locate in the inefficient region. 

Additionally, as we examine the liquidity and multifractal degree, we  confirm that 
higher liquidity ( low ILLIQ ) corresponds to smaller multifractal degree $\Delta h$. 

\begin{figure}
\centering
\includegraphics[height=9.5cm]{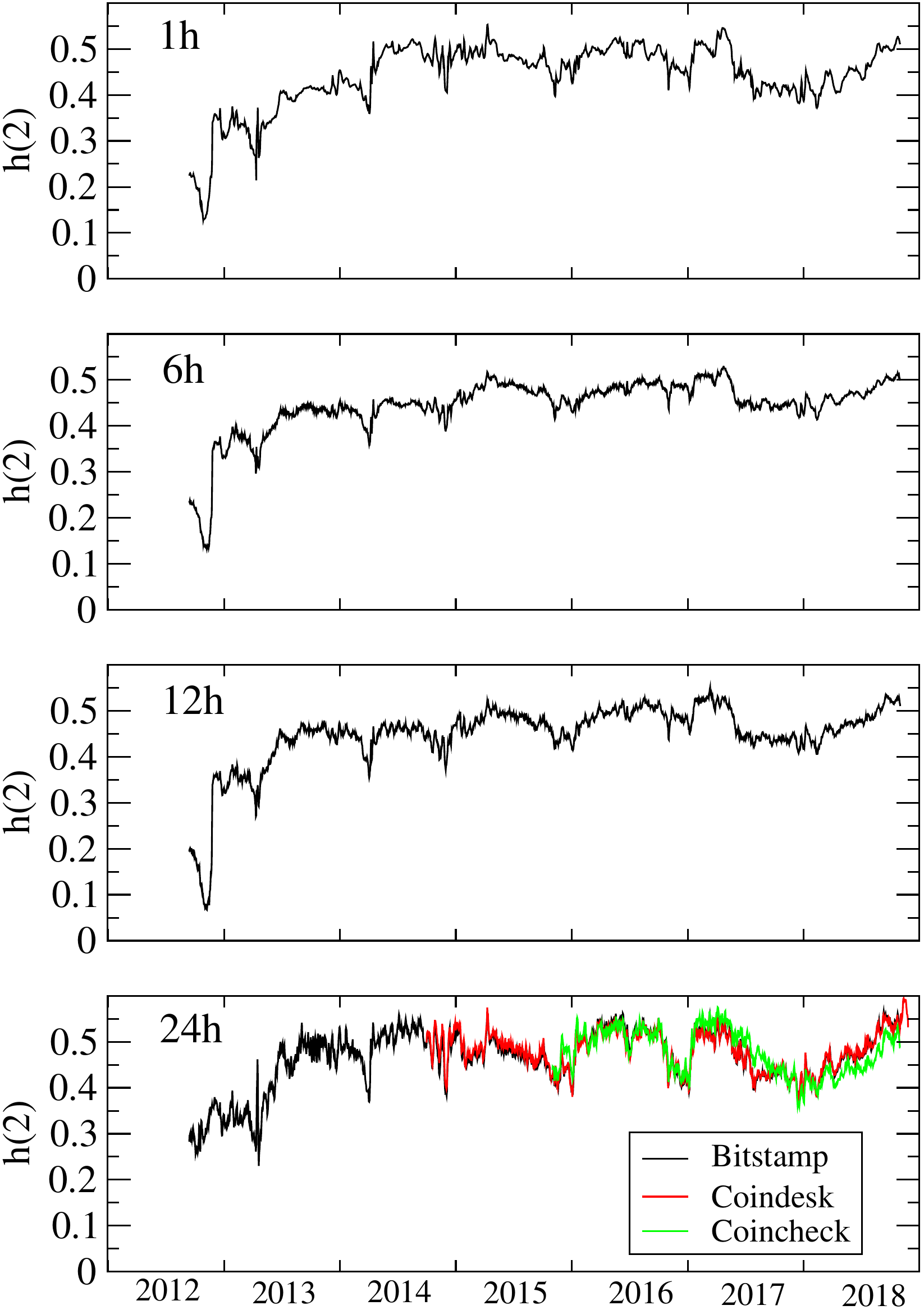}
\caption{The Hurst exponent $h(2)$ obtained from returns at $\Delta t= 1,2,6,12,24h$. 
}
\end{figure}

\begin{figure}
\centering
\includegraphics[height=9.5cm]{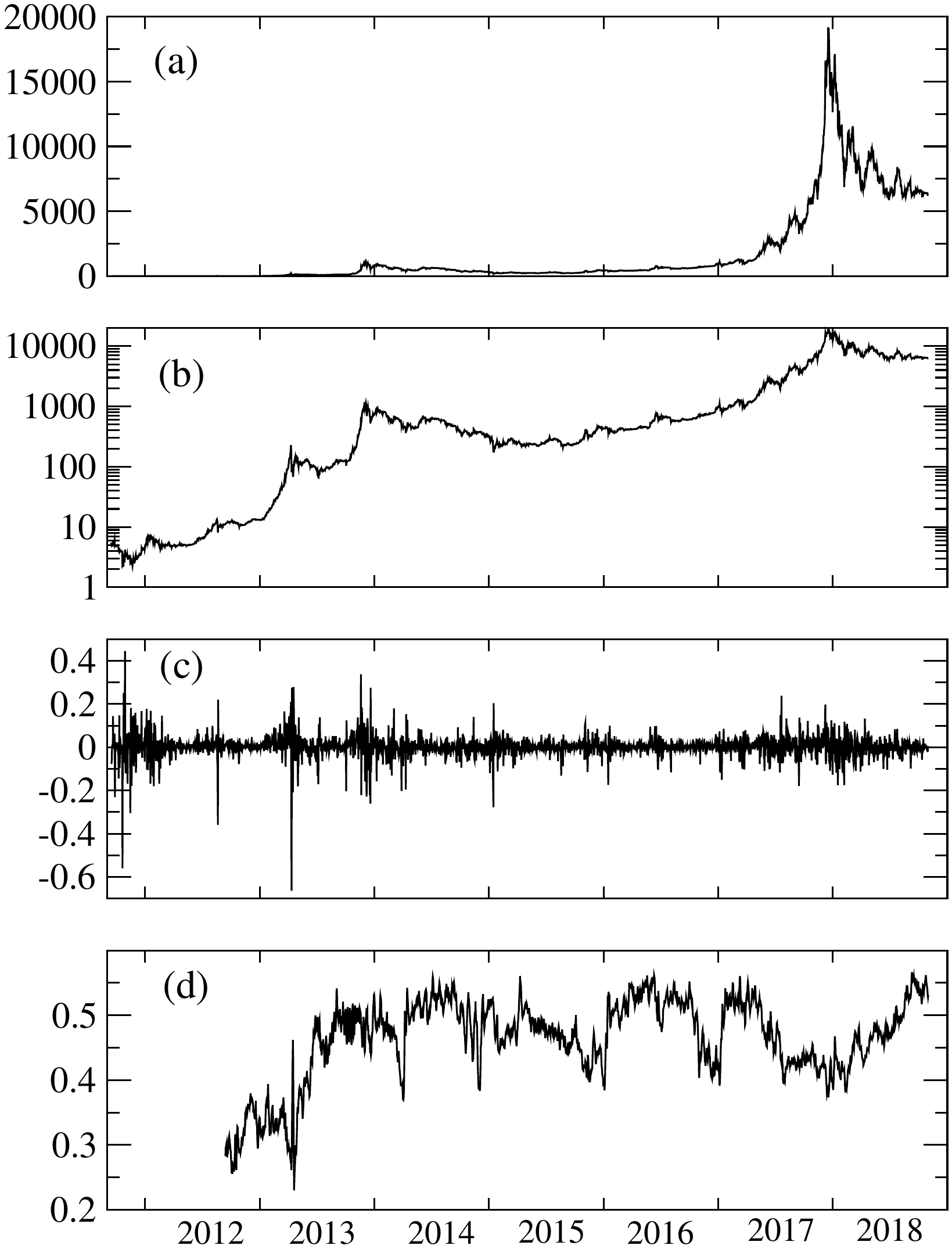}
\caption{(a) Bitcoin price history, (b) Bitcoin price history ( in logs ) (c) Return (d) Hurst exponent $h(2)$ from daily returns.
}
\end{figure}

\begin{figure}
\centering
\includegraphics[height=4.5cm]{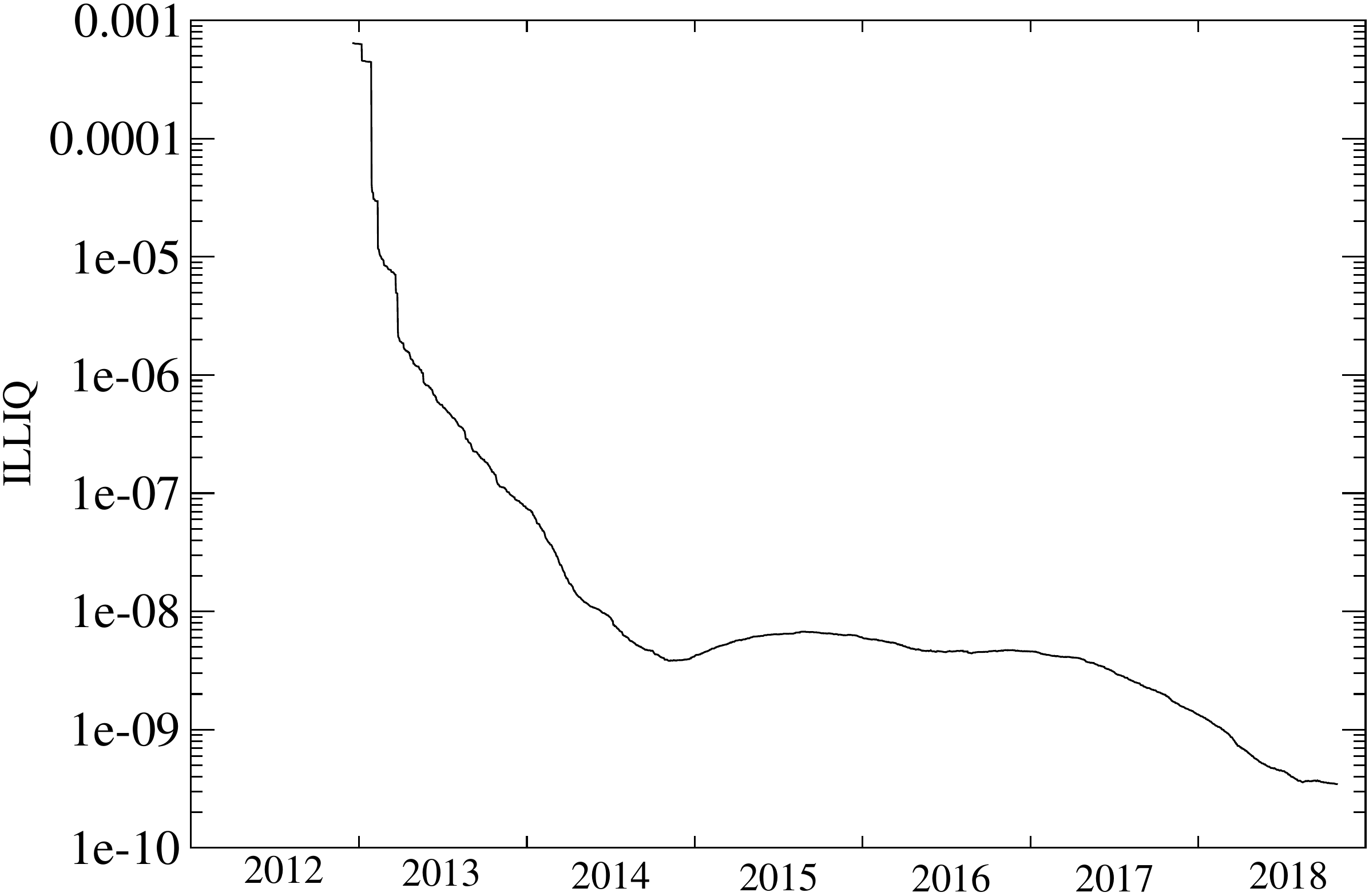}
\caption{
Amihud illiquidity measure (ILLIQ) as a function of time.
}
\end{figure}

\begin{figure}
\centering
\includegraphics[height=3.8cm]{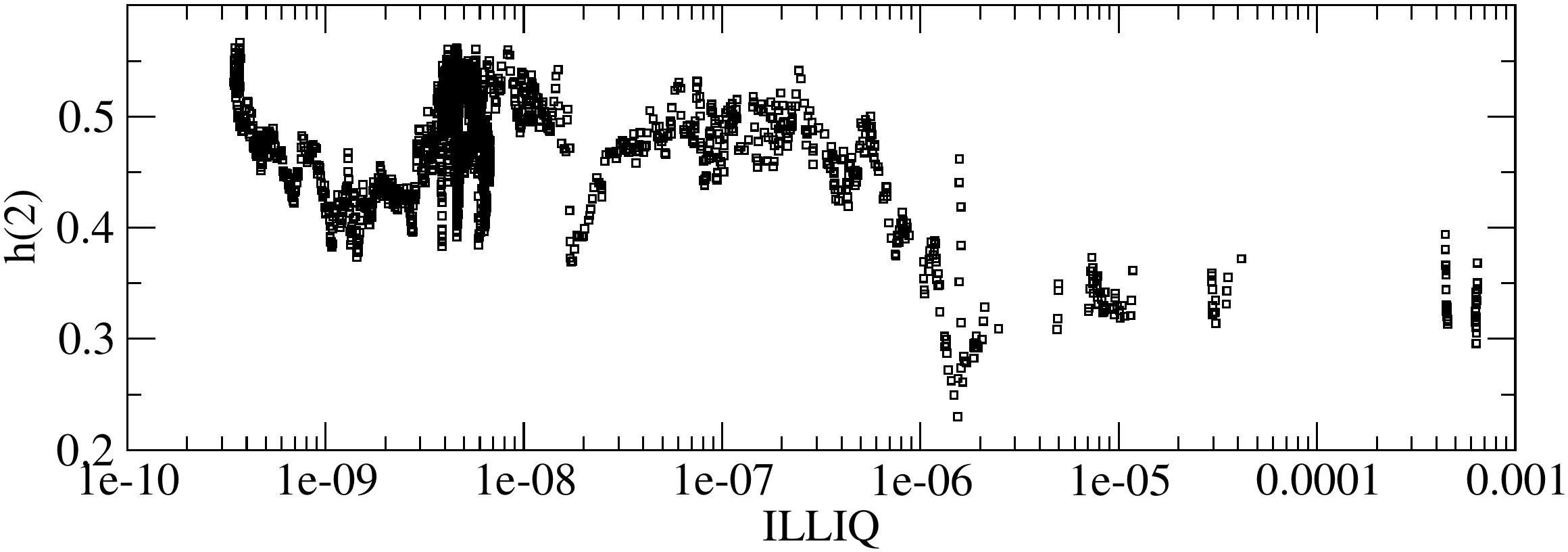}
\caption{
Hurst exponent $h(2)$ versus Amihud illiquidity measure (ILLIQ) for $\Delta t=$ 24 hours.
}
\end{figure}

\begin{figure}
\centering
\includegraphics[height=3.8cm]{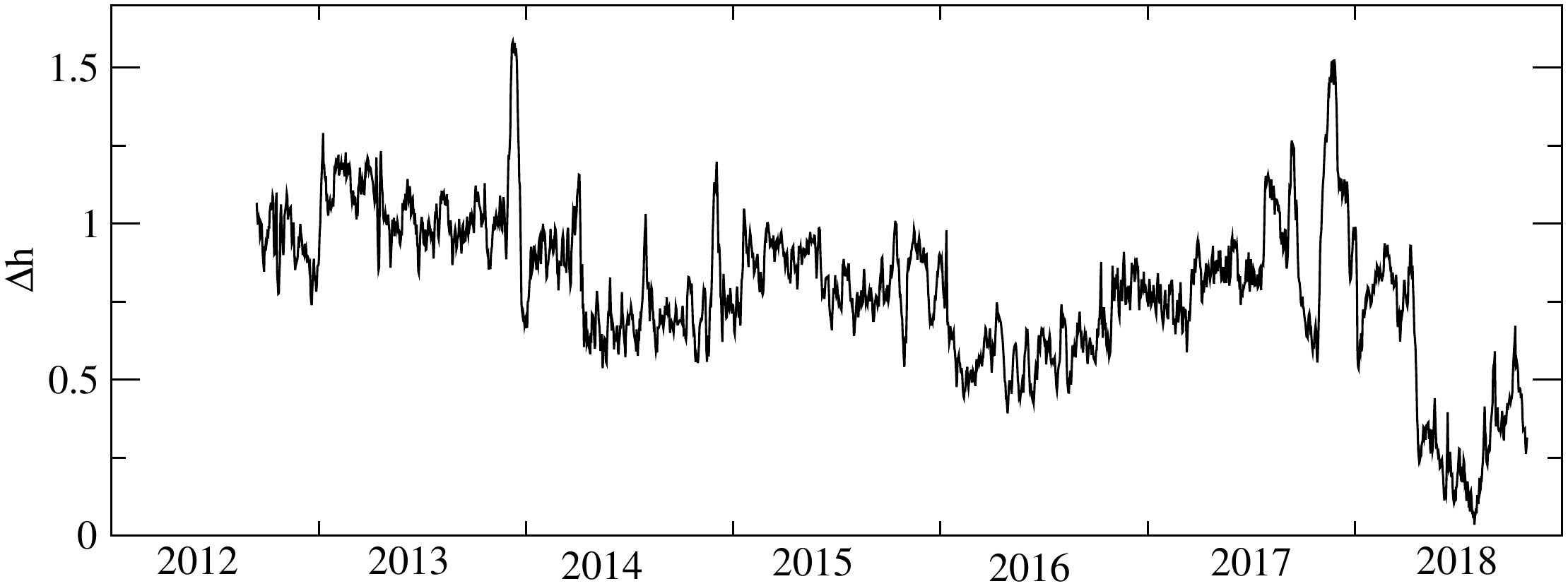}
\caption{
Multifractal degree $\Delta h$ for returns at $\Delta t= $ 24 hours as a function of time.
}
\end{figure}

\begin{figure}
\centering
\includegraphics[height=5.cm]{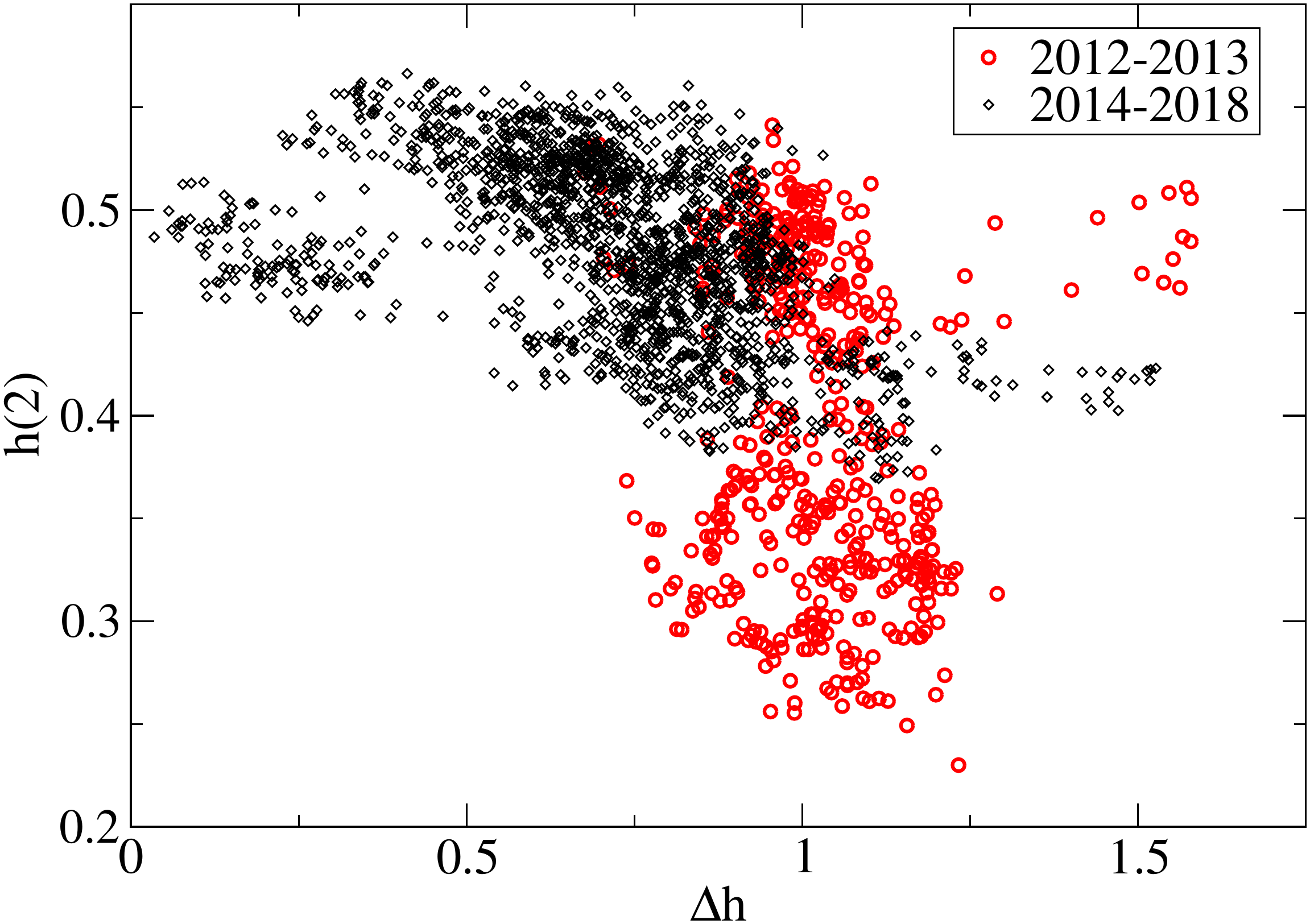}
\caption{
Scatterplots of Hurst exponent $h(2)$ and multifractal degree $\Delta h$ for $\Delta t=$ 24 hours. 
}
\end{figure}

\section{Conclusions}

We examine the relationship between market efficiency and liquidity of Bitcoin dynamically
and find that, before 2013, the liquidity is low and the Hurst exponent is significantly less than 0.5.
After 2013, liquidity increased and the Hurst exponent rose to approximately 0.5, thus 
improving market efficiency. 
For some periods, however, we observe that the Hurst exponent became significantly less than 0.5.
Since no strong persistence is seen and several periods in fact exhibit anti-persistent behavior, the overall Hurst exponent could be less than 0.5, also consistent with the results by \citet{urquhart2016inefficiency}.
Our results also support the time-varying behavior of the Hurst exponent found by \citet{alvarez2018long}.

The multifractal degree is investigated by means of the generalized Hurst exponent and
we find that the multifractal degree is related to
market efficiency in a non-linear manner.
When the multifractal degree is smaller than a threshold value,
the Hurst exponent takes on a value of approximately 0.5 and no strong correlation is observed between the multifractal degree and the Hurst exponent.
On the other hand, beyond this threshold value the Hurst exponent quickly declines  below 0.5
and the time series becomes inefficient.
This finding may indicate that even if the Hurst exponent is close to 0.5, the multifractality can appear, which implies
that the Bitcoin time series has a richer time structure that is not captured by the Hurst exponent only.

As future work, it would be interesting to investigate whether our findings are robust for other cryptocurrencies.

\section*{Acknowledgment}
Numerical calculations for this work were carried out at the
Yukawa Institute Computer Facility and at the facilities of the Institute of Statistical Mathematics.
This work was supported by JSPS KAKENHI Grant Number JP18K01556.

\section*{References}

\bibliography{mybibfile}

\end{document}